# Astronomy with Cutting-Edge ICT: From Transients in the Sky to Data over the Continents (India-US)

**Ashish Mahabal[1],\*, Ajit Kembhavi[2], Roy Williams[1] and Sharmad Navelkar[2]**

1 California Institute of Technology (Caltech), MC 249-17, 1200 E California Blvd., Pasadena CA 91125, USA
2 Inter-University Centre for Astronomy and Astrophysics (IUCAA), Ganeshkhind, Pune, 411007, India

E-Mails: aam@astro.caltech.edu; akk@iucaa.ernet.in; roy@caltech.edu; sharmad@iucaa.ernet.in

\* Author to whom correspondence should be addressed; Tel.: +1-626-395-4201; Fax: +1-626-568-9352

**Abstract:** Astronomy has always been at the forefront of information technology, moving from the era of photographic plates, to digital snapshots and now to digital movies of the sky. This has brought about a data explosion with multi-terabyte surveys already happening and upcoming petabyte scale surveys. By scanning the sky repeatedly and automatically, astronomers find rapidly changing phenomena – transients - of a great variety. Surveys like the Catalina Real-time Transient Survey (CRTS) publish details on the transients right away since many of these fade in a matter of minutes and it is important to get additional observations in order to determine their nature. This involves being able to combine a variety of datasets, small and large, in real-time. With networks like the Asia Pacific Advanced Network (APAN) and India's National Knowledge Network (NKN) we are in the realm where such a data transfer is possible in real time across continents. Here we describe the live demonstration we were able to carry out at data transfer speeds of several hundred megabits per second (Mbps) between California Institute of Technology (Caltech, USA) and the Inter-University Centre for Astronomy and Astrophysics (IUCAA, India). This project illustrates how machines can make rapid decisions in response to complex, heterogeneous data, using sophisticated software and networking. While the broader impact covers all aspects of society (disaster response, power grids, earthquakes, and many more), we have used astronomy to show how the APAN and NKN make this possible.

## 1. Introduction

Time domain astronomy has rapidly emerged as an exciting area of research in astronomy, and it will continue to grow as bigger surveys using various wavelengths come online over the next few years. The transients found touch upon a number of important scientific directions, ranging from exploration of the Solar System through supernovae to active galaxies to cosmology. It is an area that was out of reach for astronomers until recently. Since many of the transients fade quickly (order of seconds to minutes), it is critical to use other resources for follow-up and determine the nature of these objects as best as possible. This in turn involves the hard problem of choosing in real time a small number of interesting transients from a large number of well-understood ones. This necessitates advance computing and often the fast movement of bulk data. From the current rates of few to ~100 transients we are looking at rates of a million transients per night once Large Synoptic Survey Telescope (LSST), Square Kilometer Array (SKA) etc. are functional. With its various surveys Caltech is well placed to detect transients and is a Virtual Astronomy Observatory (VAO; http://www.usvao.org/) partner institute. In particular, the transients from CRTS (Drake et al., 2009) are made public in real-time and have been for the last couple of years. IUCAA, heads the Virtual Observatory in India (VOI; http://vo.iucaa.ernet.in/~voi), and is well placed to receive these events and attendant data, carry out various complex follow-up computing tasks and publish the condensed summary so that further appropriate observations can be carried out for better iterative classification (Mahabal et al., 2008) and knowledge extraction. While an optical survey like LSST will need small sets of data per transient, and will generate thousands of transients per minute, gravitational wave detectors like Laser Interferometer Gravitational Wave Observatory (LIGO) and the proposed Indian Initiative in Gravitational Wave Observations (IndIGO) will need large area searches of existing datasets for better understanding of the nature of each transient.

This process needs a number of components to come together (a) finding the transients, (b) a publishing mechanism to inform the world in real time about such events, (c) transfer of large and small auxiliary data that inform various decisions related to the follow-up as well as coordination of annotator programs that can help decipher the nature of the transients, and (d) networks that make the transfer possible. Here we briefly describe all these and concentrate mainly on the set-up and details related to the live demonstration of Gigabytes of data transfer between Caltech and IUCAA at a few hundred Mbps in response to transients.

## 2. Components

Let us look at the components mentioned in Sec. 1.

- **Telescopes/observatories and their transient streams**:
    - CRTS (**http://crts.caltech.edu**) uses data from two telescopes in Arizona and one in Australia to look for transients, i.e. astronomical objects that change significantly in brightness over a short period of time. The discoveries are announced in real-time. Activity associated with Cataclysmic Variables (CVs), Supernovae (SNe), blazars etc. causes many of the transient events observed. See Fig. 1 for an example of a few transients from CRTS.
    - Astrosat (**http://meghnad.iucaa.ernet.in/~astrosat/**) is a satellite to be launched from India in 2012. It will host 5 instruments including an all-sky monitor that will be discovering X-ray transients and publishing them. This capability is critical infrastructure for supporting gravitational wave detection.
    - IndIGO (**http://www.gw-indigo.org/tiki-index.php**) is a gravitational wave initiative that will offer a new window to the extreme physics of black holes, neutron stars, and other extreme physics. However, source detections are not well localized in the sky, requiring a search of very large amounts of data to find the corresponding transient. See Fig. 2 for a simulated example of a LIGO transient.

Since only one of these (CRTS) is functional currently, and during the time of the demonstration it was unlikely to detect transients (wrong time of the day for the telescope to be open), we made a mock stream made of transients which were announced at random but were frequent enough so that the real-time response could be seen.

- **Dissemination of events**:

SkyAlert (**http://www.skyalert.org**; Williams et al., 2009) collects and distributes astronomical events in near-real time. Each event belongs to a stream of events that come from a common source, with a common vocabulary of parameters for each event. One can browse event streams and the events themselves. One can also set up "alerts" to subselect events one finds interesting. This returns an Atom feed of the events that pass the selection.

A SkyAlert server was setup at IUCAA to receive information about these transients from the streams. Humans, machines, and telescopes can receive these events. Specific follow-up tasks can be set up in an automated fashion in response to events received e.g. get follow-up observations, retrieve older data, cross-check against older lists etc. The installation itself has a series of programs that are run automatically in response to the events. These are called annotators since their results are joined to the original transient detection information, thus forming a "portfolio" of data for each transient. All information moves as structured (XML) packets, and hence used for either presentation to the human eye, or as the basis of machine decisions. In particular, automated classification engines can run, update the data portfolio about the transient, decisions made for further actions, such as fetching more data, alerting a human, automated follow-up observation from a robotic telescope.

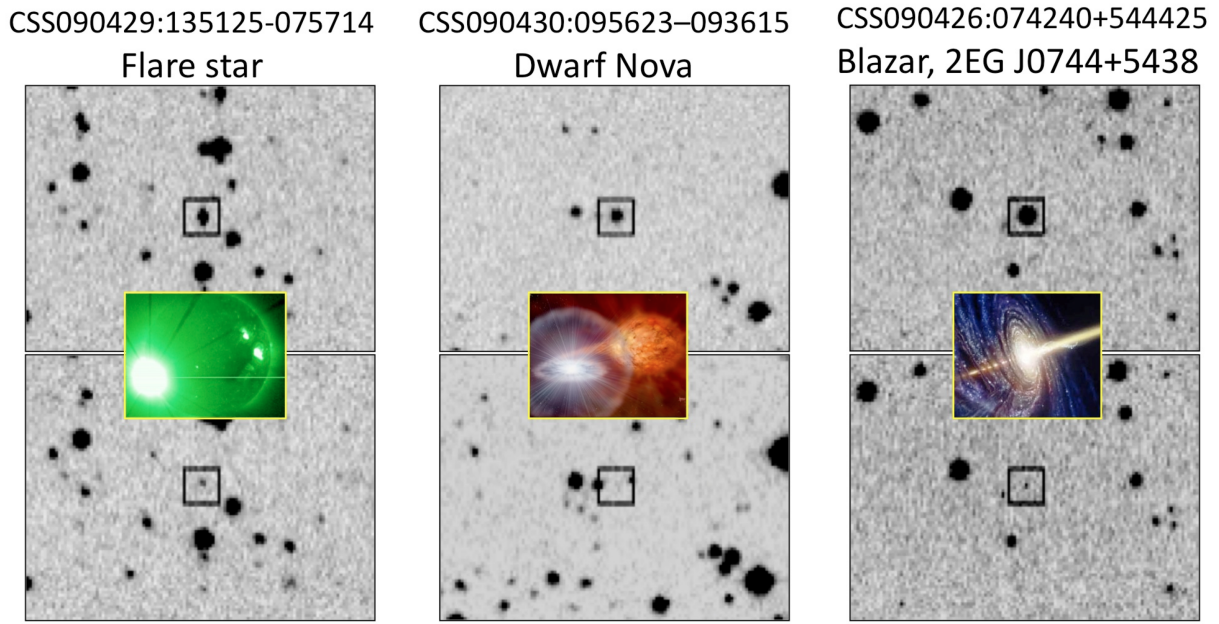

**Figure 1. Examples of a few transients from CRTS. Just the discovery images do not provide enough information for classification. Software, networking, and rapid follow-up is critical for that purpose. (Djorgovski, et al. (2011)).**

The SkyAlert server at IUCAA has the following configuration:

- Machine Type: Server HP/ML350 HP/ ML 350 Server INTEL XEON two processor quad core tower configuration
- Memory: 10 GB
- OS: CentOS 5.4 64 bit
- OS disk (SAS): 72 GB (Raid 1+0)
- Data disk (SAS): 2 TB (Raid 5)

- **The Network**

The main enabling component of this whole exercise is the fast network connecting the endpoints, Caltech and IUCAA. It is composed of a few different segments. The signal goes from Caltech in Pasadena to Los Angeles, to Tokyo, to Singapore, to Mumbai, and on to IUCAA in Pune. A WAN connection provided by the HEP network was used at Caltech (http://www.ultralight.org). The Mumbai to IUCAA part is through NKN (http://nkn.in/), a network that will eventually connect 1500 nodes allover India. The capacity is in multiples of 10 Gbps and participating institutes can connect to the network using 100 Mbps/1 Gbps speeds. The APAN (http://www.apan.net/) segments pertinent to this study have a 1 Gbps capacity. We carried out several tests prior to the live demonstration. During the demonstration 75% of the 1 Gbps bandwidth was reserved for the meeting. Only a small component (just the display) was sent to New Delhi, and as a result no significant bandwidth was required for that leg. We did not have to fine tune the different connecting points of the network at least partly due to the transfer tool, Fast Data Transfer (FDT), that we used (described next).

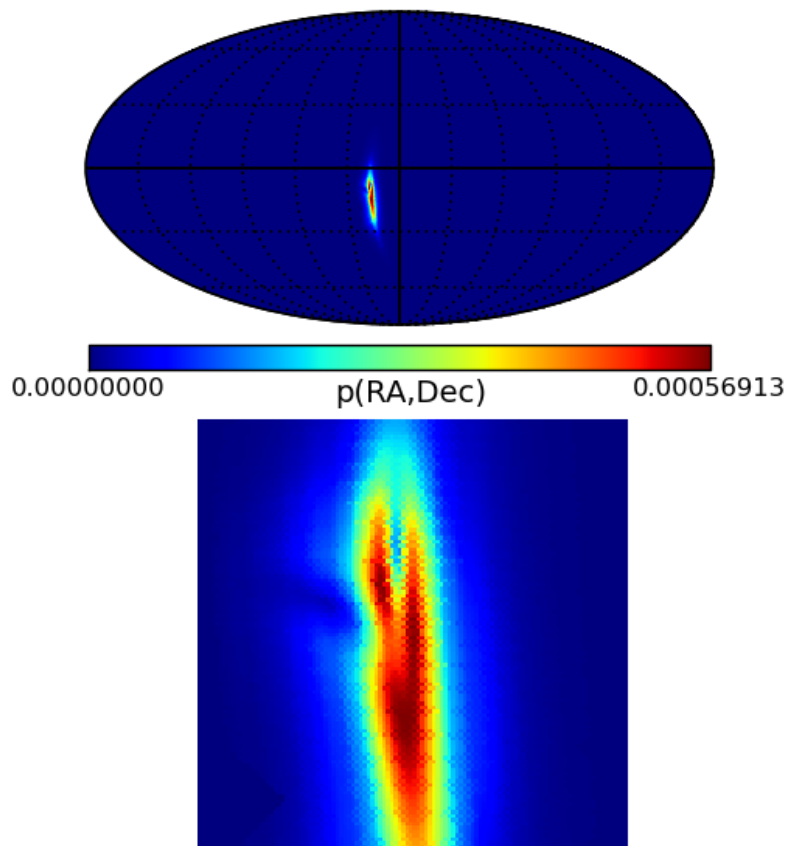

Figure 2. Top: A typical gravitational wave trigger. The uncertainty in the area is large. Bottom: zoomed in detail near the peak. For the purpose of the demonstration, instead of trying to transfer data related to the very large area, we used a one square degree area amounting to 2-5 gigabytes.

- **Auxiliary tools to aid the transfer**

We used the CERN tools FDT (http://monalisa.cern.ch/FDT/) for the actual data transfer and MonALISA (http://monalisa.cern.ch/monalisa.html) to monitor the rate at which data gets transferred. The FDT server is on a linux machine at Caltech called envoy5 and has the following configuration:

- Machine Type: 2 x AMD Dual-Core Opteron 2216 processors
- Memory: 4GB
- OS: RHEL 5.6
- OS disk: Areca Raid controller. RAID6 consisting of 16 x 1TB Hitachi drives
- Data disk: Same as OS

- **The Data**

While a number of varied data sets are used for actual transients for demonstration purposes we used tarred files of a few GB each corresponding to ~200 discrete points on the sky, amounting to a total volume of ~one TB. In particular, each of the individual sets covered one square degree area and corresponded to declinations of {-30, -20, -10, 10, 20, 30} degrees and Right Ascension of {0, 10, 20, …, 350} degrees. The sets were made of fits images from Near Earth Asteroid Tracking (NEAT; http://neat.jpl.nasa.gov/) and CRTS data. Since both these surveys avoid the center of our galaxy as it is too crowded with astronomical objects, some of the files from the potential 6*36=216 positions are not present. These datasets were also stored on envoy5.

### 3. The details

The FDT server was invoked on envoy5 with 20 parallel streams using the following command: *java -jar ~/fdt.jar -P 20.* On pavo at IUCAA a loopevents.pl script was run. This produced transients at random locations in the sky at an interval of 200+200*rand() seconds leading to a new transient every 200-400 seconds. An annotator would then determine the nearest one square degree area to pull over. Another FDT process was run in client mode on pavo embedded in the annotator called imagefetcher.py. The dataset would get pulled in one to two minutes and registered in the portfolio of the transient. The whole process would get repeated in another few minutes for a different transient. The portfolio included the discovery images as well as a link to the location of the files fetched on to pavo. The MonALISA display showed in real-time the progress of the transfer (see Fig. 3). During the demo typical rates of 550 Mbps were achieved. Several files were successfully transferred during the demo in response to the transients generated during that period.

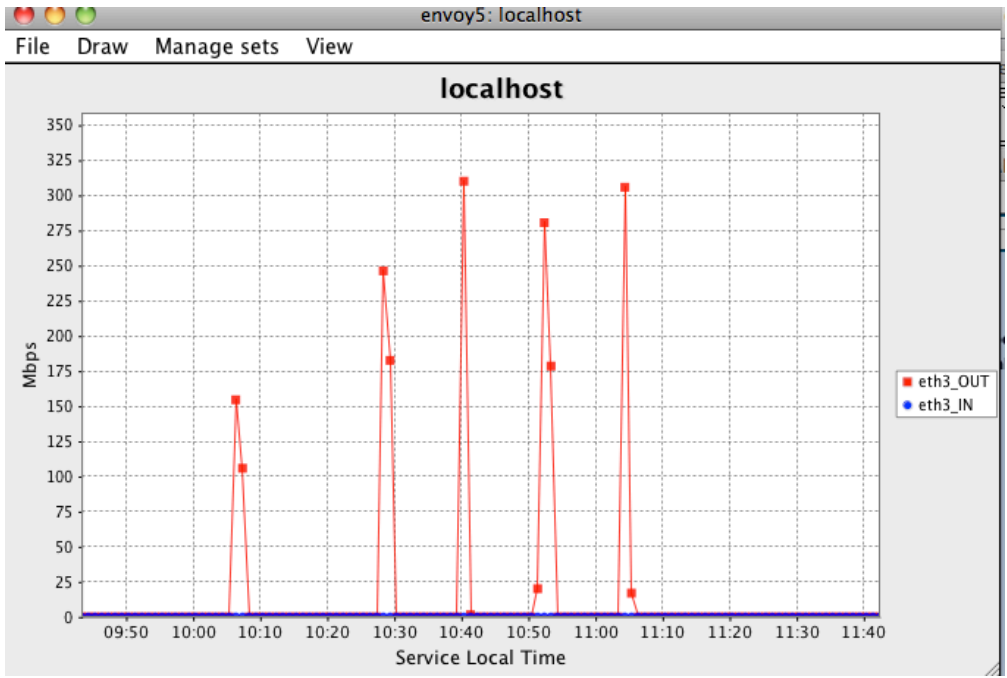

**Figure 3. During testing, when the bandwidth was not reserved, we routinely got transfer rates of 300 Mbps. During the demo we were able to reach transfer speeds of 550 Mbps.**

**4. The Future and Broader Prospects**

We have successfully demonstrated that datasets can be transferred in response to real-time triggers at high bandwidths across continents. We used only a small number of annotators to make the point since most of the annotators run locally and provide crucial functionality like classification to determine the nature of the transients. These include various programs already developed and more under development at VO-I and VAO. The need for such data transfers in astronomy is real even today and is definitely going to intensify over the next few years. Networks like APAN and NKN will thus be crucial for the advancement of transient astronomy.

We have described here transient events in astronomy where we gather data from many places, many formats, many meanings, and combine it into a meaningful portfolio, from which machines and humans can make decisions and initiate actions. In many other areas of science and technology, there are "events" and their unfolding aftermath e.g. earthquake early warning, air-traffic control, maintaining a power grid, and many other fields. When the sensors detect something out of range, we must react quickly and intelligently, based on all available data. The sensors, archives, and people are all distributed, and we want classification from different classifiers to be combined, their relevance evaluated rigorously, for fast decisions to be made with costing, and for decisions to be made in the presence of missing data. The mechanism described here is at the core of such a system. For a person there is a rich evolving web presentation of relevant data. For a machine the alert rules and annotation allow more important events to be promoted, so resources are used on these to delve deeper, run follow-ups, and to pick out what is the most relevant data in building classification. Our mathematical research enables us - in general, not just in astronomy - to rigorously combine independent classifiers to make more informed and more reliable classifications and decisions, form greater intelligence, and to pick out what is the most relevant data in building that classification.

**Acknowledgements**

**A large number of people have contributed to the success of this demonstration in direct and indirect ways. Most notably Avalon Johnson for the network setup at Caltech, and Azher Mughal for providing backend network support. Chris Biwer, Larry Price and Ajith Parameswaran helped with LIGO, IndIGO transient aspects; Sagar Shah and Neelam Bhujbal provided backend support at IUCAA; BK Gairola, PS Dhekne, RS Mani helped with NKN setup; Peter Nugent provided the NEAT data; Andrew Drake provided background programs; Dipankar Bhattacharya, Sharon Brunett, Julian Bunn, George Djorgovski, George McLaughlin, Harvey Newman, Ninan Sajeeth Philip, Sarah Ponrathnam, Dipak Singh provided high level support and encouragement. AAM acknowledges a partial support from the NSF grants AST-0834235 and AST-0909182, and the NASA grant 08-AISR08-0085. Part of this work has been facilitated by NSF grants OCI-0915473.**